\pdfoutput=1
 \documentclass[aps, prapplied,twocolumn,showpacs,superscriptaddress,preprintnumbers, longbibliography,nofootinbib]{revtex4-2}
\usepackage[normalem]{ulem}
\usepackage{amsmath}
\usepackage{amssymb}
\usepackage{physics}  %
\usepackage{epsfig}

\usepackage{url} 

\usepackage{graphicx}
\usepackage{hyperref}
\usepackage{tabu}
\usepackage{boldline}
\usepackage{xspace}
\usepackage{slashed}
\usepackage{multirow}
\usepackage{booktabs}
\usepackage{diagbox}
\usepackage{tabularx}
\usepackage{comment}
\usepackage{floatrow}
\newfloatcommand{capbtabbox}{table}[][\FBwidth]
\usepackage{color}
\usepackage[normal]{subfigure}
\usepackage{enumitem}
\usepackage[utf8]{inputenc}
\usepackage{colortbl}
\usepackage{upgreek}
\usepackage{lineno}
\usepackage[separate-uncertainty=true]{siunitx}
\usepackage[dvipsnames]{xcolor}
\AtBeginDocument{\RenewCommandCopy\qty\SI}

\newcommand\sjt{\bgroup\markoverwith{\textcolor{blue}{\rule[0.5ex]{2pt}{0.4pt}}}\ULon}

\DeclareCaptionJustification{justified}{\justified}

\newcommand{\myrange}[3]{[\num{#1},~\num{#2}]~\unit{#3}}  

\DeclareSIUnit{\electron}{e\textsuperscript{-}}
\DeclareSIUnit{\bar}{bar}
\DeclareSIUnit{\pix}{\textup{pixel}}
\DeclareSIUnit{\day}{\textup{day}}
\DeclareSIUnit{\epd}{\electron\per\pix\per\day}
\DeclareSIUnit{\ADU}{\text{A.D.U.}}

\usepackage{lipsum} 

\begin{document}

\preprint{}

\title{Charge Trap Analysis in a SENSEI Skipper-CCD: Understanding Low-Energy Backgrounds in Rare-Event Searches}


\author{Agustin Brusco}
\email{agustin.brusco@gmail.com}
\affiliation{\normalsize\it 
Universidad de Buenos Aires, Facultad de Ciencias Exactas y Naturales, Departamento de Física, Buenos Aires, Argentina}

\author{Bruno Sivilotti}
\email{brunosivilotti@hotmail.com}
\affiliation{\normalsize\it 
Universidad de Buenos Aires, Facultad de Ciencias Exactas y Naturales, Departamento de Física, Buenos Aires, Argentina}

\author{Ana M. Botti}
\affiliation{\normalsize\it 
Fermi National Accelerator Laboratory, PO Box 500, Batavia IL, 60510, USA}
\affiliation{\normalsize\it Kavli Institute for Cosmological Physics, University of Chicago, Chicago, IL 60637, USA}

\author{Brenda Cervantes}
\affiliation{\normalsize\it 
Fermi National Accelerator Laboratory, PO Box 500, Batavia IL, 60510, USA}

\author{Ansh Desai}
\affiliation{\normalsize\it 
Department of Physics and Institute for Fundamental Science, University of Oregon, Eugene, Oregon 97403, USA}

\author{Rouven Essig}
\affiliation{\normalsize\it 
C.N.~Yang Institute for Theoretical Physics, Stony Brook University, Stony Brook, NY 11794, USA}

 \author{Juan Estrada}
\affiliation{\normalsize\it 
Fermi National Accelerator Laboratory, PO Box 500, Batavia IL, 60510, USA}

\author{Erez Etzion}
\affiliation{\normalsize\it 
 School of Physics and Astronomy, Tel-Aviv University, Tel-Aviv 69978, Israel}
 
\author{Guillermo Fernandez Moroni}
\affiliation{\normalsize\it 
Fermi National Accelerator Laboratory, PO Box 500, Batavia IL, 60510, USA}

\author{Stephen E. Holland}
\affiliation{\normalsize\it Lawrence Berkeley National Laboratory, One Cyclotron Road, Berkeley, California 94720, USA}

\author{Ian Lawson}
\affiliation{\normalsize\it SNOLAB, Lively, ON P3Y 1N2, Canada}

\author{Steffon Luoma}
\affiliation{\normalsize\it SNOLAB, Lively, ON P3Y 1N2, Canada}

\author{Santiago E. Perez}
\affiliation{\normalsize\it 
Fermi National Accelerator Laboratory, PO Box 500, Batavia IL, 60510, USA}
\affiliation{\normalsize\it 
Universidad de Buenos Aires, Facultad de Ciencias Exactas y Naturales, Departamento de Física, Buenos Aires, Argentina}
\affiliation{\normalsize\it 
CONICET - Universidad de Buenos Aires, Instituto de Física de Buenos Aires (IFIBA). Buenos Aires, Argentina}

\author{Dario Rodrigues}
\affiliation{\normalsize\it 
Universidad de Buenos Aires, Facultad de Ciencias Exactas y Naturales, Departamento de Física, Buenos Aires, Argentina}
\affiliation{\normalsize\it 
CONICET - Universidad de Buenos Aires, Instituto de Física de Buenos Aires (IFIBA). Buenos Aires, Argentina}

\author{Javier Tiffenberg}
\affiliation{\normalsize\it 
Fermi National Accelerator Laboratory, PO Box 500, Batavia IL, 60510, USA}
\affiliation{\normalsize\it 
Universidad de Buenos Aires, Facultad de Ciencias Exactas y Naturales, Departamento de Física, Buenos Aires, Argentina}

\author{Sho Uemura}
\affiliation{\normalsize\it 
Fermi National Accelerator Laboratory, PO Box 500, Batavia IL, 60510, USA}

\author{Yikai Wu}
\affiliation{\normalsize\it 
C.N.~Yang Institute for Theoretical Physics, Stony Brook University, Stony Brook, NY 11794, USA}
\affiliation{\normalsize\it 
Department of Physics and Astronomy, Stony Brook University, Stony Brook, NY 11794, USA} 

\begin{abstract}
\noindent Skipper Charge-Coupled Devices (Skipper-CCDs) are ultra-low-threshold detectors capable of detecting energy deposits in silicon at the eV scale. Increasingly used in rare-event searches, one of the major challenges in these experiments is mitigating low-energy backgrounds. In this work, we present results on trap characterization in a silicon Skipper-CCD produced in the same fabrication run as the SENSEI experiment at SNOLAB. Lattice defects contribute to backgrounds in rare-event searches through single-electron charge trapping. To investigate this, we employ the charge-pumping technique at different temperatures to identify dipoles produced by traps in the CCD channel. We fully characterize a fraction of these traps and use this information to extrapolate their contribution to the single-electron background in SENSEI. We find that this subpopulation of traps does not contribute significantly but more work is needed to assess the impact of the traps that can not be characterized. 
\end{abstract}

\maketitle

\section{Traps and single-electron backgrounds}\label{sec:intro}

Rare-event searches through electron recoils in silicon with Skipper Charge Couple Devices (Skipper-CCDs)~\cite{Smith2010, Holland:2003, Janesick1990,Wen1974, Tiffenberg:2017aac} have made significant progress since the first result published by the SENSEI collaboration~\cite{Crisler:2018gci}. 
The low-energy threshold of these sensors (1.12~eV) and their deep sub-electron resolution enable the detection of particle interactions that ionize as few as a single electron.

After demonstrating Skipper-CCDs' potential for light-dark-matter detection, new efforts emerged to increase the experimental sensitivity by scaling the detector mass while mitigating the different backgrounds~\cite{Abramoff:2019dfb, Sensei2020, senseicollaboration2023sensei,DAMIC-M:2023gxo,aguilar2022oscura}. 
Skipper-CCDs' unprecedented precision in identifying, characterizing, and reducing single-electron backgrounds has recently led to the lowest dark current ever achieved in any silicon device or ultraviolet-to-near-infrared photodetector~\cite{sensei1e}. Understanding the origin of the remaining dark events is crucial for rare-event searches, as these background sources may arise from intrinsic processes within the sensor~\cite{senseiSEE,ampLight} or from environmental backgrounds~\cite{,Du_2022,Du:2023soy,Sensei2020}.

Recently, the Oscura Collaboration demonstrated that contaminants or defects in the silicon lattice within the Skipper-CCD buried channel can trap charge and release it at a later time, partially explaining single-electron backgrounds~\cite{oscurasensors, trapsOscura}. Charges from high-energy interactions may become trapped and later released, either shortly after, contributing to charge transfer inefficiency, or much later, adding to the intrinsic dark current. In their study, the Oscura Collaboration used a Skipper-CCD fabricated by Microchip Inc. to implement a pocket-pumping technique and published a protocol to identify charge traps based on their physical location and intensity.

In this work, we present results for a Skipper-CCD designed at LBNL and fabricated at Teledyne DALSA in the same wafer as CCDs used in SENSEI at SNOLAB~\cite{senseicollaboration2023sensei}. The sensor was packaged at Fermilab and operated at the LAMBDA laboratory at the University of Buenos Aires. In Section~\ref{sec:section2}, we describe the experimental setup and protocol used for data acquisition. In Section~\ref{sec:identification} we describe the trap identification technique and in Section~\ref{sec:analsys} the trap energy and temperature dependence. Finally, we summarize and discuss the impact of the trap density on SENSEI dark current in Section~\ref{sec:discussion}.

\section{Pocket-Pumping measurements}\label{sec:section2}

The pocket-pumping technique implemented in this work involves uniformly illuminating the CCD with an external light source, followed by repeatedly shifting the resulting charge carriers, holes in this case, back and forth within the three pixel phases. If a trap exists in one of the edge phases, charges may be captured and released at a later time spilling into a neighbor and creating a signal dipole between the pixels; a detailed description of this method is presented in~\cite{trapsOscura}. The number of charge carriers trapped and released to the neighbor pixel, or intensity of the dipole ($\mathrm{I(t_{ph})}$), is calculated as half of the difference in charge between the two pixels.
The intensity depends on the number of {\it{pump}} cycles ($\mathrm{N_{pump}}$) and the duration of each clock state ($\mathrm{t_{ph}}$) as defined later in Eq.~\ref{I_fit}.

We used a high-resistivity silicon Skipper-CCD with $1658\,\times\,572$\,pixels, each of $(15 \times 15\,)\upmu\mathrm{m}^2$, housed inside a vacuum chamber equipped with a vacuum pump and a cryocooler. The Skipper-CCD has a readout amplifier at each corner, enabling the parallel readout of each quadrant. A custom proportional–integral–derivative (PID)  temperature controller designed at LAMBDA~\cite{Pietra2022} allowed us to control the sensor temperature between \qty{120}{\kelvin} and \qty{200}{\kelvin}. The chamber wall facing the CCD featured a window to enable external illumination. We controlled light exposure using a white OLED screen (generic SSD1306 display) positioned in front of the window, with optical diffusers to enhance uniformity. An external module, designed to be coupled to the vacuum chamber, housed the screen, diffusers, and control electronics while shielding the CCD from environmental light. We used a Raspberry Pi pico to control the OLED screen, and a Low-Threshold Acquisition board (LTA)~\cite{LTA} to operate and read the CCD.

\begin{figure}[t]
\includegraphics[trim={0.0cm 0.0cm 0.0cm 0.0cm},clip,width=1\textwidth]{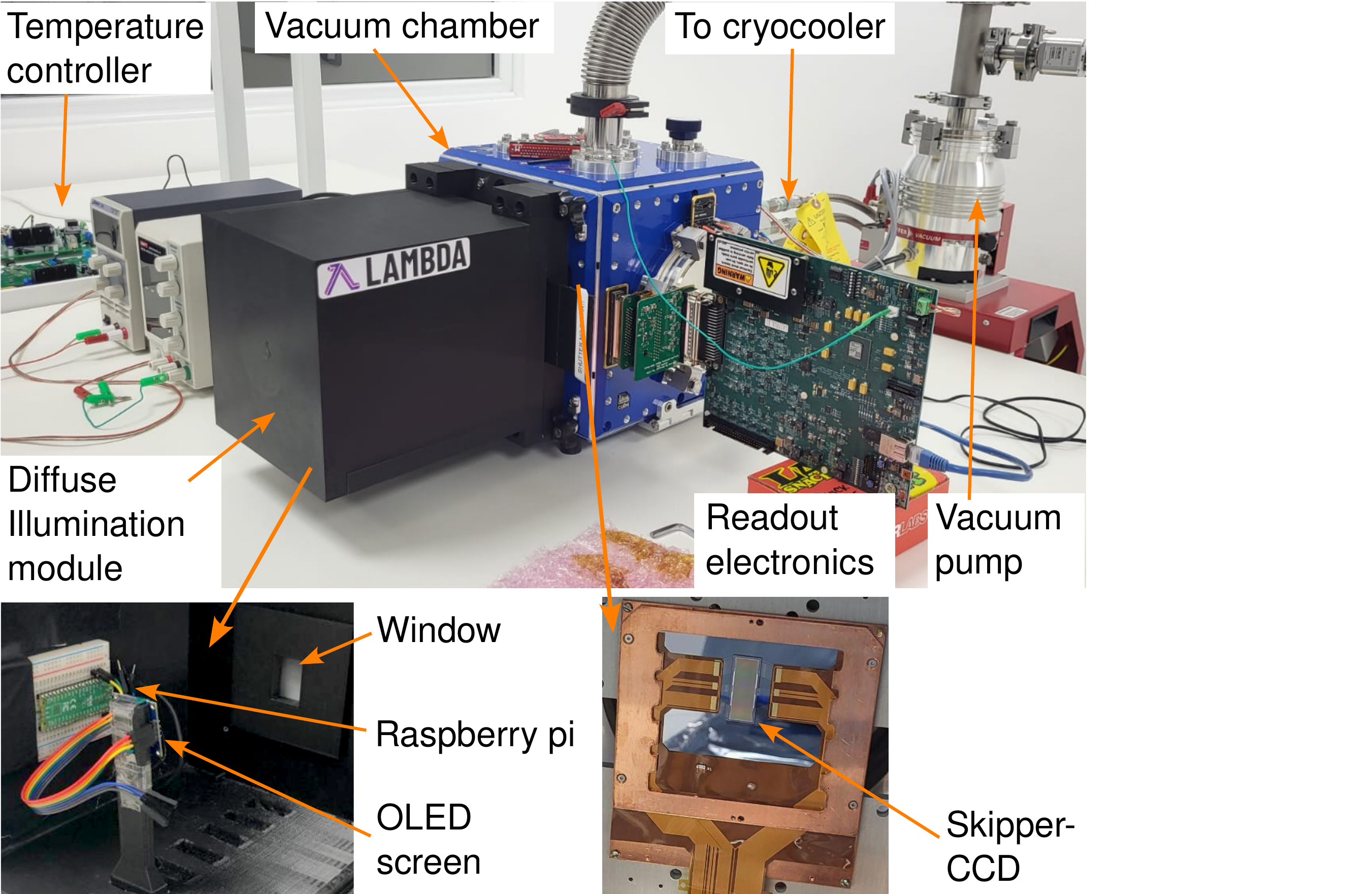}
\caption{The Skipper-CCD is mounted in a copper tray that is thermally coupled to a cold finger connected to a cryocooler. The readout electronics is provided by an LTA board. A resistive heater, controlled through a PID feedback loop, allows operation at different temperatures. For uniform illumination, we use an external black-box module that houses an OLED screen controlled by a Raspberry Pi. A diffuser is placed in front of an opening aligned with a fused-silica window on the vacuum vessel, allowing light to reach the Skipper-CCD.}\label{fig:setp}
\end{figure}

Each measurement involved setting the system temperature, illuminating the sensor's active area with the OLED screen to achieve an average charge of approximately \qty{4000}{\electron} per pixel, and executing the Pocket-Pumping protocol with $\mathrm{N_{pump}}=$\num{3000} to obtain well-defined, non-saturated dipoles. The CCD was read out using a single Skipper sample ($\mathrm{N_{samp}}=$\num{1}), to enhance readout speed and minimize environmental background interactions, such as cosmic rays. For each temperature, \num{25} images were acquired while varying $t_{ph}$ between \qty{3.3}{\us} and \qty{1.37}{\ms}. This process was repeated for 15 different temperatures in the range of \myrange{126}{195}{\kelvin}, resulting in a total of 375 images.

\section{Trap identification}
\label{sec:identification}

\begin{figure}[t]
\includegraphics[trim={0.0cm 0.0cm 0.0cm 0.0cm},clip,width=1\textwidth]{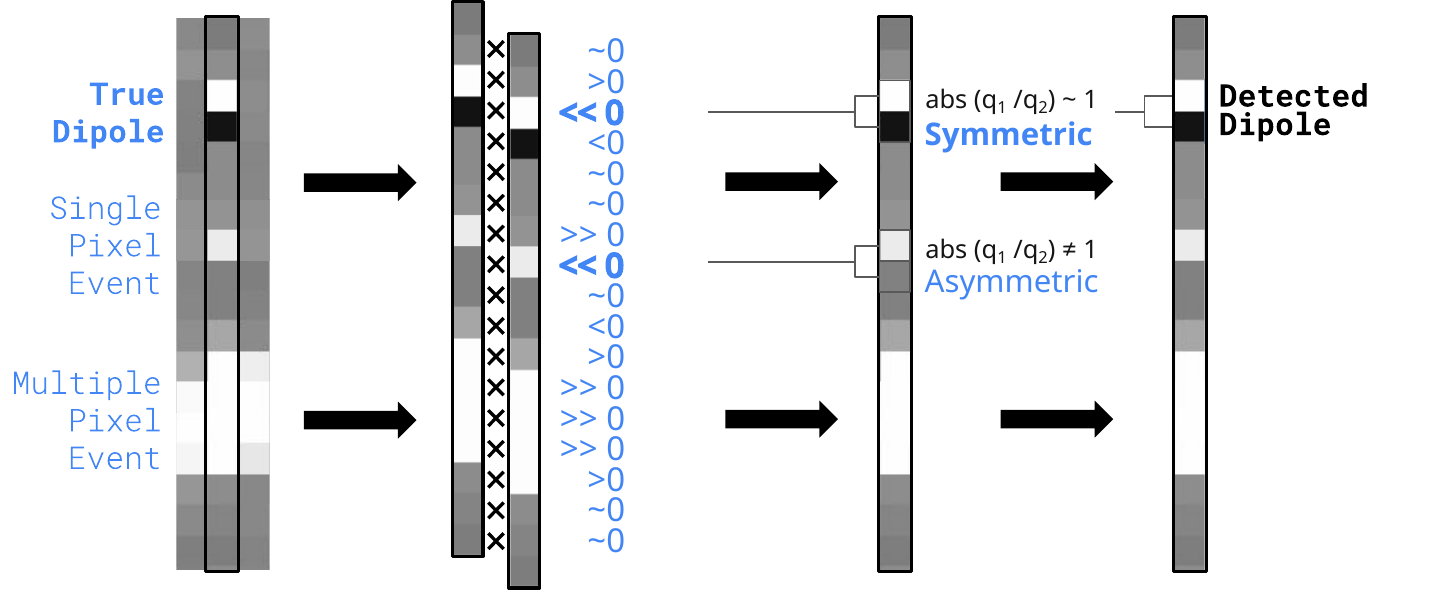}
\caption{Diagram of the dipole detection algorithm. From left to right: the computation of the self-correlation, the filtering by a threshold value $C$, and the symmetry filter before confirming a detection. $q_{1}$ and $q_{2}$ denote the charges measured in the two adjacent pixels along the pumping direction. }\label{fig:dipole_algorithm}
\end{figure}

We searched for dipoles in the images to identify traps in the Skipper-CCD. The process began by subtracting the median charge value for each row. Then, we applied the algorithm illustrated in Fig.~\ref{fig:dipole_algorithm}, which multiplies the charge values of neighboring pixels within the same column. If a dipole is present, this computation yields a negative value. In contrast, if only charge carriers from the illumination are present, the result is approximately zero. Multi-pixel events caused by high-energy backgrounds produce a positive value. To locate the dipoles, we searched for multiplications resulting in negative values below a threshold, $C$. Since traps generate dipoles where the pixel charge deviates by more than $3\sigma$ from the mean charge of the image, we fixed $C=-(3\sigma)^2$.

Furthermore, due to charge conservation, dipoles are expected to be symmetric: within fluctuations, their negative and positive pixels (after median subtraction) should exhibit equal absolute charge values. To exploit this property, we implemented a symmetry filter that selects neighboring pixels whose absolute charge values differ by less than 30\%, defined by the condition $ 0.7 < q_{1}/q_{2} <1.3$, where $q_{1}$ and $q_{2}$ denote the charges measured in the two adjacent pixels along the pumping direction.

Finally, we selected only dipoles that satisfied the selection criteria in more than two images acquired at the same temperature but with different $\mathrm{t_{ph}}$ values. Once we identified a dipole location associated with a trap in the CCD, we searched for it in all other images taken at the same temperature, obtaining its intensity for each $\mathrm{t_{ph}}$.

The dipole intensity depends on $\mathrm{N_{pump}}$, $\mathrm{t_{ph}}$, the trap depth ($\mathrm{D_t}$), the probability of capturing a charge carrier ($\mathrm{P_c}$), and the probability of emitting it ($\mathrm{P_e}$)~\cite{trapsOscura}. In this approximation, $\mathrm{P_c}$ acts as a scaling factor~\cite{HallPc} that varies with temperature, while $\mathrm{P_e}$ can be expressed as a function of $\mathrm{t_{ph}}$, and the emission characteristic time ($\mathrm{\tau_e(T)}$), namely:

\begin{align}\label{I_fit}
    \mathrm{I(t_{ph}) =N_{pumps}D_t P_c \left(e^{-\frac{t_{ph}}{\tau_e}}-e^{-8\frac{t_{ph}}{\tau_e}}\right)}
\end{align}

\begin{figure}[t]
\includegraphics[trim={0.0cm 0.0cm 0.0cm 0.0cm},clip,width=1\textwidth]{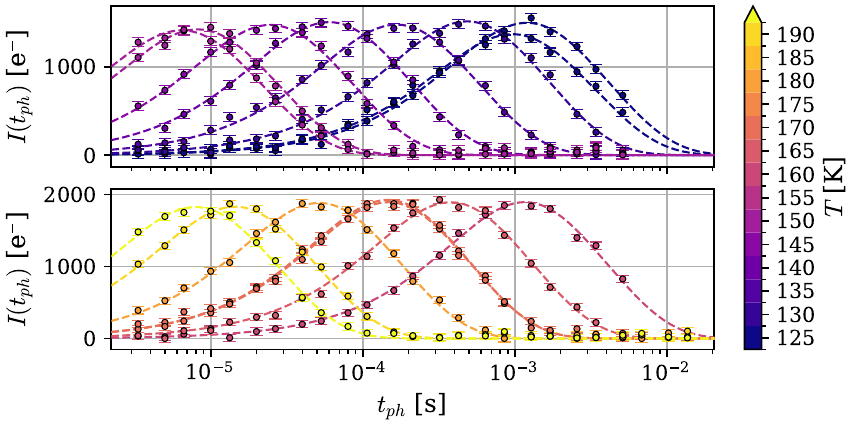}
\caption{ 
Measured dipole intensity as a function of $t_{ph}$ for two detected traps: one appearing at low temperatures (top) and another detectable at high temperatures (bottom). Data are shown for several temperatures, along with their corresponding fits using Eq.~\eqref{I_fit}.}\label{fig:I(tph)}
\end{figure}

Each image provides a single data point for the $\mathrm{I(t_{ph})}$~vs.~$\mathrm{t_{ph}}$ curve for a selected temperature and dipole, as plotted in Fig.~\ref{fig:I(tph)}. We show $\mathrm{I(t_{ph})}$ curves for two traps (top and bottom panels) at different temperatures (color scale), along with fits using Eq.~\eqref{I_fit}. We performed a $\chi^{2}$ goodness of fit test selecting the curves with p-value\,$>$\,\num{0.05} to reject pixel pairs identified as dipoles that did not behave as described by the model; 138 traps passed the test and were included in the subsequent analysis. 
As shown in Fig.~\ref{fig:I(tph)}, the curves shift leftward as temperature increases and are well described by the fitted model.

\begin{figure}[t]
\includegraphics[trim={0.0cm 0.0cm 0.0cm 0.0cm},clip,width=1\textwidth]{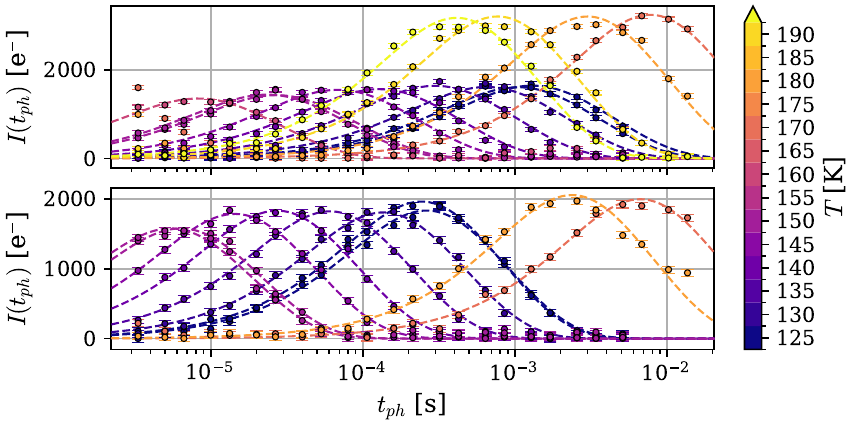}
\caption{
Measured dipole intensity as a function of $t_{ph}$ for two detected traps (top and bottom) that showed two distinct responses at low temperatures ($<150$K) and again at high temperature $>170$K). Data are shown for several temperatures, along with their corresponding fits using Eq.~\eqref{I_fit}.
}\label{fig:two_resp}
\end{figure}

In addition to these 138 well-characterized traps, we identified five traps that matched the model's predicted behavior (Eq.~\eqref{I_fit}) in two separate temperature ranges. We display two examples of these dual-response traps in the top and bottom panels of Fig.~\ref{fig:two_resp}, showing that below $\sim$150\,K the intensity curve shifts leftward as the temperature increases, as expected. In addition to this, for temperatures above 170\,K, another curve appears on the right, which also shifts leftward with rising temperatures and follows the shape described by Eq.~\eqref{I_fit}. This phenomenon suggests the presence of defects with two distinct resonances or, less likely, the appearance of two different traps under the same pixel. These five dual-response dipoles were excluded from the subsequent analysis.

\begin{figure}[t]
\includegraphics[trim={0.0cm 0.0cm 0.0cm 0.0cm},clip,width=1\textwidth]{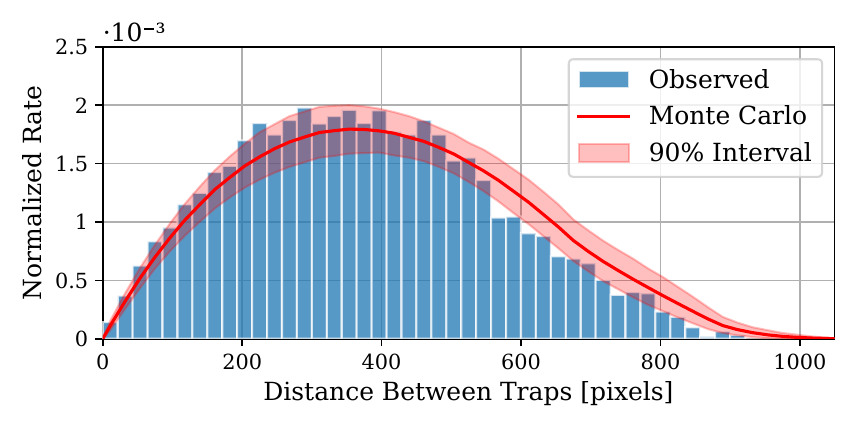}
\caption{Distribution of the distance between identified traps on the CCD. We also show the mean and the 90\% interval of toy  Monte Carlo simulations assuming traps are uniformly distributed}\label{fig:spatial_dist}
\end{figure}

To analyze the spatial distribution of the 138 selected traps, we performed 10$^4$ toy Monte Carlo simulations, modeling the traps under the assumption of spatial uniformity to use this as a benchmark for comparison.
Fig.~\ref{fig:spatial_dist} shows the histogram of the distances between all identified traps, along with the mean curve from the simulations assuming a uniform spatial distribution and the corresponding 90\% confidence band. A slight deviation between the simulation and the data is observed for distances greater than 500 pixels, indicating that traps in the data are closer together than expected from a uniform distribution. This is expected since traps may arise from spatially localized phenomena, such as fabrication defects or hits from high-energy particles.

\section{Trap temperature dependence and energy}\label{sec:analsys}

\noindent For the selected dipoles, Eq.~\eqref{I_fit} allows us to extract $\mathrm{\tau_e}$, which is a function of the lattice temperature (T), the trap energy ($\mathrm{E_t}$) and cross section ($\sigma$), the charge carriers' thermal velocity ($\mathrm{v_{th}}$), and the effective density of states in the conduction band ($\mathrm{N_c})$:
\begin{align}\label{tau(T)}
    \mathrm{\tau_e = \frac{1}{\sigma v_{th}N_c}e^{\frac{E_t}{K_B T}}},
\end{align}

\noindent where $\mathrm{K_B}$ is the Boltzmann constant, and $\mathrm{v_{th}}$ and $\mathrm{N_c}$ are also temperature-dependent, defined as:

\begin{align}
    \mathrm{v_{th} = \sqrt{\frac{3K_B T}{m_{cond}}}~~~and~~~
    N_c = 2 \left[ 2\pi m_{dens} \frac{K_B T}{h^2}\right]^{3/2}}.
\end{align}

\noindent here, h is the Plank's constant, $\mathrm{m_{cond} \simeq 0.41\,m_e}$ and $\mathrm{m_{dens} \simeq 0.94\,m_e}$ are the holes effective mass for conductivity and density of states respectively between 100 and 200\,K~\cite{GreenCon}, and $\mathrm{m_e}$ is the electron mass at rest.

\begin{figure}[t]
\includegraphics[trim={0.0cm 0.0cm 0.0cm 0.0cm},clip,width=1\textwidth]{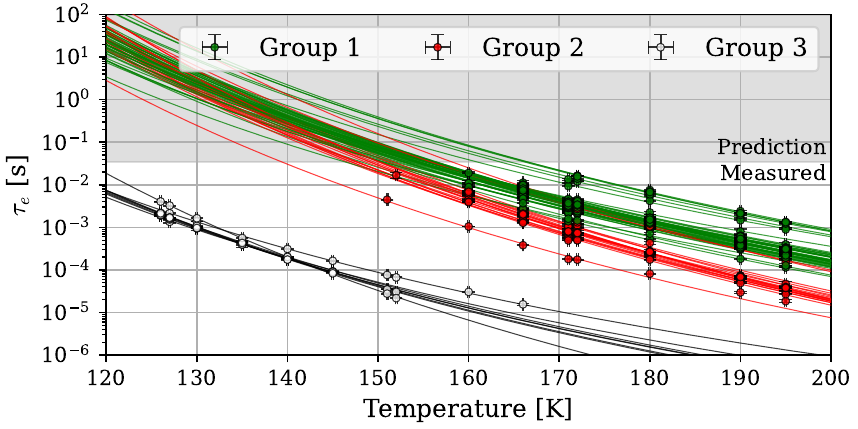}\\
\vspace*{0.1in}
\hspace*{0.02in}\includegraphics[trim={0.0cm 0.0cm 0.0cm 0.0cm},clip,width=1.0\textwidth]{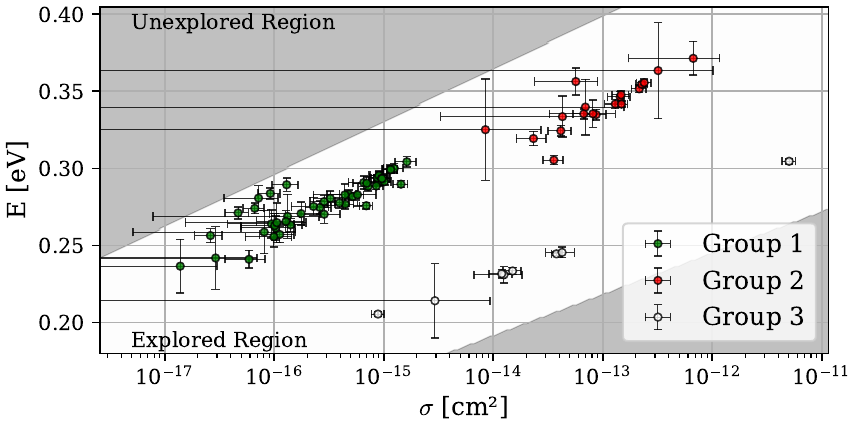}
\caption{Parameters fitted from trap data. Colors indicate the trap type according to its intrinsic characteristics. Gray shadows indicate the region in the parameter space unattainable given the experimental conditions. Measurements of the $\tau_e$ versus the temperature of a given trap fitted with Eq.~\eqref{tau(T)}~(top). Distribution of energies and cross sections of the characterized traps on the CCD~(bottom).}\label{fig:tau_T}
\end{figure}

We then analyzed the emission characteristic time of the traps with respect to the temperature. Among the 138 selected traps, only 77 were observed for at least four temperatures. 
For these, the top panel of Fig.~\ref{fig:tau_T} presents the values of $\tau_e(T)$ obtained after fitting the dipole intensity as a function of $t_{ph}$ for all observed temperatures (i.e. each one of the curves shown in Fig.~\ref{fig:I(tph)} contributes a single data point). 
Using Eq.~\eqref{tau(T)}, we can fit the four (or more) data points for each dipole in this plot (Fig.~\ref{fig:tau_T}~top) and extract the trap energy E and cross-section $\sigma$. The gray region indicates the experimentally inaccessible range of $\tau_e$ for our current setup since longer times require longer measurements, which we cannot perform on the surface due to environmental background. However, we can extrapolate the measurements to lower temperatures using Eq.~\eqref{tau(T)} to predict the emission characteristic time of all traps at about 130\,K, the typical operating temperatures of underground Skipper-CCD experiments.

In the bottom panel of Fig.~\ref{fig:tau_T} we present the trap energy and cross section fitted using Eq.~\eqref{tau(T)}. The traps are grouped into three distinct populations according to their energy and capture cross section. The spread observed in the points corresponding to a single trap population across different cross sections may arise from the fact that each trap can be located anywhere within the pixel and at varying positions relative to the pumping phase, which can affect the capture probability. 
A trap located near the center of the charge packet may exhibit a different capture probability than one near the edge, even if both belong to the same trap population, since local dynamics within the silicon lattice can vary~\cite{HallPc}. The observed variations could also result from the presence of different contaminants~\cite{Bilgi_traps}. Further analysis is needed to discriminate between these two possibilities and to enable the reliable identification of specific contaminant species from these measurements.
The gray shade in Fig.~\ref{fig:tau_T} indicates the region in the parameter space that was not explored given the experimental conditions.

\begin{figure}[t]
\includegraphics[trim={0.0cm 0.0cm 0.0cm 0.0cm},clip,width=1\textwidth]{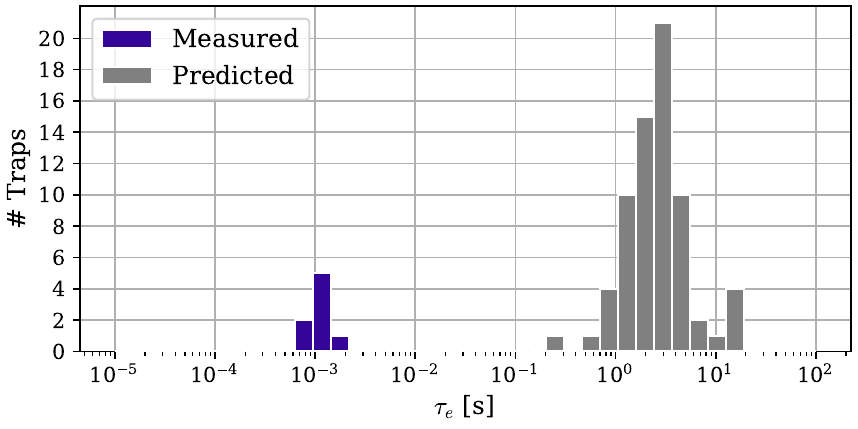}
\caption{Emission characteristic times of CCD traps at 130~K. Shown are traps directly measured at 130~K, with characteristic times of approximately \qty{1}{\ms}, and a second population extrapolated from traps characterized at higher temperatures (as illustrated in the top panel of Fig.~\ref{fig:tau_T}), with characteristic times around \qty{1}{\s}.
}\label{fig:spectrum}
\end{figure}

In Fig.~\ref{fig:spectrum}, we show the $\tau_e$ distribution extracted from the top panel of Fig.~\ref{fig:tau_T} for CCD traps at 130~K, the typical operating temperature of Skipper-CCDs used in dark-matter searches. Two distinct groups are evident: one directly measured at 130~K, with characteristic times of approximately \qty{1}{\ms}, and another, extrapolated from traps characterized at higher temperatures, with characteristic times around \qty{1}{\s}. The impact of these traps was evaluated using the toy Monte Carlo simulation developed in Ref.~\cite{trapsOscura}.

To estimate how traps contribute to the single-electron rate measured by SENSEI during the 2020 run, this toy Monte Carlo generates a uniform spatial distribution of defects and assigns each a decay time determined by the CCD operating temperature, trap energy, and capture cross section. The simulation then processes existing SENSEI images, extracting high-energy events ($>20,e^-$) and producing a new synthetic image in which the single-electron rate includes both exposure-dependent and exposure-independent components generated using the values in~\cite{Sensei2020}.

The readout of high-energy events is modeled by shifting charge packets toward the serial register—the CCD structure where charge is transferred pixel by pixel to the readout stage for sequential measurement~\cite{janesick2001scientific}. If a charge packet passes through a pixel containing a trap, the capture process is simulated such that a single electron may subsequently be released either in the same pixel or in a later one. Charge carriers can also be re-captured as they propagate toward the readout amplifier.

From this procedure, two sets of images are produced: one including the effects of traps and one without them. The same data analysis pipeline used by SENSEI is then applied to verify whether the initially simulated single-electron rate can be accurately recovered. 
We find that electrons generated by deferred charge from these trap species are completely rejected by the standard SENSEI masks applied during data analysis. Although these masks were originally designed to remove events arising from other sources—such as charge-transfer inefficiencies and Cherenkov photons produced by high-energy particles—they also prove highly effective at rejecting events caused by deferred charge release. This efficiency stems from the fact that such releases typically occur near high-energy events, consistent with the measured characteristic times. As a benchmark, if no masking were applied, the contribution of the traps measured in this work to the SENSEI dataset would amount to approximately $0.003~e^-/\text{pix}/\text{day}$.

In future work, we plan to extend this study to include traps with longer characteristic times, which may produce events uncorrelated with high-energy tracks.

\section{Summary and outlook}\label{sec:discussion}

In this work, we have characterized charge traps in a Skipper-CCD fabricated in the same run as the sensors deployed in the SENSEI experiment at SNOLAB. Using the pocket-pumping technique over a broad range of temperatures, we identified and studied 138 traps, extracting their emission time constants and classifying them by their trap energy and capture cross section. These parameters were used to model the trap behavior and extrapolate their influence at the typical operating temperature of SENSEI detectors ($\sim$130K).

Our measurements show that most of the characterized traps have emission times either near 1ms or 1s at 130 K. We simulated their impact using a toy Monte Carlo framework to assess their contribution to the single-electron background observed in SENSEI~\cite{Sensei2020}. We find that, under current masking strategies~\cite{sensei2020Sup}, the deferred charge generated by these traps is effectively suppressed in the analysis pipeline.

Importantly, the full trap population in these devices may include species with longer emission times or lower capture probabilities, which are inaccessible with the current surface-level measurements. Therefore, additional measurements—either at lower temperatures or in low-background underground environments—will be necessary to determine the contribution of slower traps and to fully assess the role of lattice defects in shaping the low-energy background.

These results are a key step toward understanding the intrinsic contributions to the single-electron background in Skipper-CCDs. Continued trap characterization across different fabrication batches and under various environmental conditions will be essential for reducing backgrounds in future low-threshold dark matter searches such as Oscura and DAMIC-M. Ultimately, improving our understanding of charge trapping mechanisms will enhance the sensitivity of these experiments and help unlock new regions of parameter space in the search for rare interactions.

\begin{acknowledgments}
We are grateful for the support of the Heising-Simons Foundation under Grant No.~79921. This document was prepared by the SENSEI collaboration using the resources of the Fermi National Accelerator Laboratory (Fermilab), a U.S. Department of Energy, Office of Science, Office of High Energy Physics HEP User Facility. Fermilab is managed by Fermi Research Alliance, LLC (FRA), acting under Contract No.~DE-AC02-07CH11359.
The CCD development work was supported in part by the Director, Office of Science, of the DOE under No.~DE-AC02-05CH11231.
The U.S. Government retains and the publisher, by accepting the article for publication, acknowledges that the U.S. Government retains a non-exclusive, paid-up, irrevocable, world-wide license to publish or reproduce the published form of this manuscript, or allow others to do so, for U.S. Government purposes.
\end{acknowledgments}

\bibliography{references.bib}

\end{document}